latex file using espcrc2.sty (nuclear physics style file included)
\documentstyle[twoside,fleqn,espcrc2]{article}

\newcommand{\AmS}{{\protect\the\textfont2
  A\kern-.1667em\lower.5ex\hbox{M}\kern-.125emS}}

\title{Measurement of the penetration depth and coherence length in
U(1) and SU(2) dual Abrikosov vortices}

\author{Vandana Singh, Dana A. Browne, and Richard W. Haymaker
  \thanks{R.W.H. and V.S. are supported in part by the U. S. Department of
          Energy under grant DE-FG05-91ER40617 and D.A.B. is supported
          in part by the National Science Foundation under grant No.
          NSF-DMR-9020310} \\ {} \\
Department of Physics and Astronomy, Louisiana State University, Baton Rouge,
Louisiana, 70808,USA}

\begin{document}

\begin{abstract}
We calculate the electric field and the curl of the magnetic monopole
current for U(1) and for SU(2) in the maximal Abelian gauge in the
mid-plane between a quark antiquark pair.  The results can be understood
as a dual Abrikosov vortex in the Ginzburg-Landau theory.
\end{abstract}

\maketitle

\section{INTRODUCTION}

The most natural and likely scenario to explain the mechanism by which
QCD confines quarks is that superconducting currents act to form a flux
tube connecting a separated quark from a residual hadron.  There is
ample evidence for this in a related theory, i.e. the confining phase
of U(1) in 3+1 dimensions as can be seen for example in terms of the
`bulk properties' of the vacuum state\cite{dt}.  In a recent paper we
mapped out in considerable detail the structure of the dual Abrikosov
vortex in the presence of static quarks \cite{shb}.  As in ordinary
superconductivity, the key is to study the relationship of the curl of
the supercurrent to the electric field since this is the origin of the
Meissner effect. Supercurrents form a solenoid which squeezes field
lines into a tube.

In non-Abelian theories the situation is far less clear since analytic
techniques are intractable.  One encouraging approach is to partially
fix the gauge, leaving a U(1) gauge freedom in the maximal Abelian
gauge\cite{klsw,sy}. The U(1) Dirac monopoles that occur are abundant
in the confined phase and dilute in the unconfined phase (at finite
temperature) exactly as in U(1) gauge theory.   We present here the
first {\it direct} evidence that a dual Abrikosov vortex also forms in
a lattice simulation of pure gauge SU(2) with static quarks\cite{sbh}.

\section{U(1)}

The basic techniques for U(1) are described in more detail in
Ref. \cite{shb}.  Magnetic monopoles are the current carriers
responsible for the dual superconductivity.  These monopoles are
defined in a 3-volume by the DeGrand-Toussaint\cite{dt} construction
and it is convenient to think  of them as a link on the dual lattice,
making world lines which define a conserved current density $J_M$.  The
key idea is to measure the line integral of $J_M$ around a dual
plaquette, i.e. $\vec{\nabla}\times \vec{J}_M$.   This quantity has a
very strong correlation with a Wilson loop, giving a clean signal for
the solenoidal behavior of the currents surrounding the electric flux
between oppositely charged particles.

\begin{figure}[htb]
\begin{picture}(200,140)(0,0)
\newsavebox{\curla}
\savebox{\curla}{
\put(100,100){\line(1,0){40}}
\put(100,140){\line(1,0){40}}
\put(100,100){\line(0,1){40}}
\put(140,100){\line(0,1){40}}
\put(118,117){$\bullet$}
}
\newsavebox{\cord}
\savebox{\cord}{
\put(24,-8){z}
\put(-8,24){x}
\put(-20,-12){y}
\put(0,0){\line(1,0){24}}
\put(0,0){\line(0,0){24}}
\put(0,0){\line(-2,-1){12}}
}
\newsavebox{\curlb}
\savebox{\curlb}{
\put(100,140){\line(-2,-1){20}}
\put(100,140){\line(1,0){40}}
\put(140,140){\line(-2,-1){20}}
\put( 80,130){\line(1,0){40}}
\put(109,133){$\bullet$}
}
\newsavebox{\final}
\savebox{\final}{
\put(30,140){\usebox{\cord}}
\put(50,-2){\usebox{\curla}}
\put(50,42){\usebox{\curla}}
\put(48,-1){\usebox{\curlb}}
\put(72,11){\usebox{\curlb}}
\put(80,139){\rule{14.3mm}{0.6mm}}
\put(150,139){\rule{14.3mm}{0.5mm}}
\put(183,125){\line(0,1){42}}
\put(159,113){\line(0,1){42}}
\put(183,167){\line(-2,-1){24}}
\put(183,125){\line(-2,-1){24}}
\put(65,180){a)}
\put(130,180){b)} }
\put(-5,-80){\usebox{\final}}
\end{picture}
\caption{Operators for a) electric field b)
$\vec{\nabla}\times \vec{J}_M$ on a fixed time slice.}
\label{curl}
\end{figure}

\begin{figure}[t]
\begin{picture}(200,300)(0,0)
\end{picture}
\caption{Profile of $\vec{E}$, $\vec{\nabla}\times \vec{J}_M$ and fluxoid in
U(1).}
\label{fluxoid}
\end{figure}

Figure 1 shows the operators for the (a) electric field $\vec{E}$ and
(b) $\vec{\nabla}\times \vec{J}_M$.  The longitudinal electric field is
given by a z-t plaquette which is depicted by a bold line for fixed
time.  The corresponding operator $\vec{\nabla}\times \vec{J}_M$ is
built from four 3-volumes which appear as squares since the time
dimension is not shown.  Passing through the center of each square is
the link dual to the 3-volume.  One takes the monopole number $n$ in
each 3-volume and associates the value $n e_M$ with the corresponding
dual link. To obtain $\vec{\nabla}\times \vec{J}_M$ one performs the
discrete `line integral' around the dual plaquette. Notice from this
construction that $\vec{E}$ and $\vec{\nabla}\times \vec{J}_M$ take
values at the same location within the unit cell of the lattice,
indicated by the bold face line in Fig. 1(b).

Because of their vector nature, $\vec{E}$ and $\vec{\nabla}\times
\vec{J}_M$ average to zero in the vacuum.   We introduce static charges
by correlating these quantities with a Wilson loop in the z-t plane.
the electric field is given by $E_z= \langle W P_{zt} \rangle / (a^2 e
\langle W \rangle)$.  (Here $W$ and $P$ stand for the complex value of
the Wilson loop and plaquette, not the real part.) $\vec{\nabla}\times
\vec{J}_M$ is measured by correlating the operator shown in Fig. 1(b)
with the Wilson loop in place of $P_{zt}/e$.  Measuring $\vec{E}$ and
$\vec{\nabla}\times \vec{J}_M$ in the presence of a $3 \times 3$ Wilson
loop on the plane midway between the two charges provided profiles of
these two quantities which are shown in Fig. 2(a) and 2(b).  They were
found to satisfy the London relation\cite{London}
\begin{equation}
E_z - \frac{\lambda^2}{c} (\vec{\nabla}\times \vec{J}_M)_z =
\Phi_{e} \delta^2(\vec{x}_{\perp}).
\label{1}
\end{equation}
This is the London limit for a (dual) Abrikosov vortex containing one
unit of quantized electric flux.  The left hand side of Eqn.(1) defines
the fluxoid\cite{London}.  A one parameter fit of the data to this
relation gives a London penetration depth\cite{London} of $\lambda/a =
0.482\pm0.008$ for $\beta = 0.95$. The resulting fluxoid is shown in
Fig. 2(c). The value on axis, $1.016\pm 0.014$ agrees very well with one
unit of electric flux, $\Phi_e=1/\sqrt{\beta}=1.026$.

Using Maxwell's equations to eliminate $\vec{J}_M$ gives an equation
for $\vec{E}$ alone which has a solution
\begin{equation}
E_z(r) = \frac{\Phi_e}{2\pi \lambda^2} K_0(r/\lambda).
\label{2}
\end{equation}
This is shown as a dashed curve in Fig. 2(a). There are hence three
methods to determine $\lambda$: the cancellation of the off-axis
profiles in Fig. 2(a) and 2(b), the value of the fluxoid on axis in Fig
2(c), and the fit of the electric field in Eqn.(\ref{2}), all of which
agree.

Further evidence for the Meissner effect is given by the dependence of
$\lambda$ on the value of $\beta$.  In an ordinary superconductor, the
transition to the normal state is accompanied by a divergence in the
penetration depth.  We see analogous behavior in the transition to the
unconfined phase at $\beta\approx 1.0$.  For $\beta$ = 0.90, 0.95,
0.97, and 0.99 we find corresponding $\lambda$ values of 0.32, 0.48,
0.60, and 0.8.

\section{SU(2)}

These methods are immediately applicable to SU(2) in the maximal
Abelian gauge. In this approach one fixes the gauge in order to
identify Abelian link and coset variables.  Kronfeld {\em et
al.\/}\cite{klsw} have shown that the monopoles defined using these
Abelian link variables are indicators of confinement very much as they
are in the U(1) theory.  Further Suzuki {\em et al.\/}\cite{sy} have
shown that the static quark potential can be obtained very well from
the Abelian link variables alone.  Encouraged by these studies, we
examine the dual Meissner effect here again looking for the detailed
structure of the Abrikosov vortex.

The gauge is fixed by maximizing the quantity
\begin{equation}
R = \sum_{s,\mu} Tr
[ \sigma_3 \widetilde{U}(s,\mu)\sigma_3\widetilde{U}^{\dagger}(s,\mu)],
\label{3}
\end{equation}
where $\widetilde{U}(s,\mu) = V(s)U(s,\mu)V^{-1}(s+\mu)$. The Abelian link
variable is $\widetilde{U}(s,\mu)_{11}$ normalized to modulus 1.
Maximizing R corresponds to diagonalizing $X(s)$ at each site, where
\begin{eqnarray}
X(s) = \sum_{\mu}&&[U(s,\mu)\sigma_3U^{\dagger}(s,\mu)+\\
\nonumber
&& U(s-\mu,\mu)\sigma_3U^{\dagger}(s-\mu,\mu)].
\label{4}
\end{eqnarray}
Let $Z(s,\mu)$ be the off diagonal matrix element of $X(s)$ and
$|Z|^2$ denotes the average value of the modulus squared which is a
measure of gauge fixing.  Typically we needed about 650 sweeps to attain
$|Z|^2 \approx 10^{-5}$ on a $13^3 \times 14$ lattice for $\beta=2.4$.
Three gauge fixing methods were used: (1) a modification of Metropolis
to accept if $R$ increases and reject of $R$ decreases, (2) maximize
exactly locally at alternate sites, (3) overrelaxation in which we
chose the square of the gauge transformation of method (2) in order
to sample configurations better.  These methods were alternated in
a manner to speed up gauge fixing.

After gauge fixing the measurements proceed exactly as in the U(1)
case with one important difference.   In the U(1) case we knew
the value of the Abelian charge, $e$. We divided the plaquette by
$e$ in order to get the electric field, and we multiplied the integer
valued lattice operator for $\vec{\nabla}\times \vec{J}_M$ by
$e_M = 2 \pi / e$ to normalize the magnetic current. In this case
since we do not have a local U(1) action, we do not know the value
of $e$.  However note that the relative normalization of the two terms does
not depend on $e$ and hence we can determine the London penetration depth
as in the U(1) case. Hence our reported values of $\vec{E}$ and
$\vec{\nabla}\times \vec{J}_M$  do not include division by $e$; quantities
are given up to a common normalization factor.

\begin{figure}[htb]
\begin{picture}(200,200)(0,0)
\end{picture}
\caption{Profile of $\vec{E}$ in SU(2).}
\label{su2e}
\end{figure}

\begin{figure}[htb]
\begin{picture}(200,200)(0,0)
\end{picture}
\caption{Profile of $\vec{\nabla}\times\vec{J}_M$ in SU(2).}
\label{su2c}
\end{figure}

Figures 3 and 4 show the profile of the flux tube on the midplane
between a quark antiquark pair for $E$ and $\vec{\nabla}\times \vec{J}_M$
respectively.  These data came from 480 measurements for $\beta = 2.4$,
for a $3 \times 3$ Wilson loop on a $13^3 \times 14$ lattice with 50
updates between measurements.  These profiles are similar to U(1)
results, Figs. 2(a) and 2(b).  However, unlike the U(1) case
-$\vec{\nabla}\times\vec{J}_M$ has a positive value at one point off axis, and
there is clearly {\it no} linear combination of the off axis data for
$E$ and $\vec{\nabla}\times\vec{J}_M$ which will give zero.  However what we
are
seeing is more of the structure of the Abrikosov vortex than in the
U(1) case.  In fact the more general Ginzburg-Landau theory\cite{Tink}
does explain this behavior.  The coherence length $\xi$ is the length scale
measuring the thickness of the normal-superconducting interface.  In
U(1) we were insensitive to the coherence length since it must be
significantly smaller than a lattice spacing.  In SU(2) the coherence
length can be estimated as the radius of the node in $\vec{\nabla}\times
J_M$, i.e. $\approx 1.5 a $.

The Abrikosov vortex is a solution to the Ginzburg-Landau equations where
the GL order parameter in polar coordinates varies as
\begin{equation}
\psi = \psi_{\infty}f(r)\,e^{i\theta}
\label{5}
\end{equation}
and the radial amplitude $f(r)$ is given approximately\cite{Tink} by $f(r) =
\tanh(r/\xi)$.  The generalized fluxoid relation is
\begin{equation}
\Phi_0 = \int_S \vec{E}\cdot\vec{ds}
- \frac{\lambda^2}{c}\int_S \vec{\nabla}\times \frac{\vec{J}_M}{f^2(r)}\cdot
\vec{ds}.
\label{6}
\end{equation}
The curves in Figs. 3 and 4 represent our fit to the data.  The details
are given in Ref. \cite{sbh}.  The fitted values we obtained are
$\lambda/a = 1.05 \pm 0.12$ and $\xi/a = 1.48 \pm 0.12$.

The aim of this study was to find operators that are sensitive to the
detailed properties of the the flux tube.
$\vec{\nabla}\times\vec{J}_M$ is such a probe.  Having demonstrated the
utility of measuring this quantity, one must of course go to larger
lattices and larger Wilson loops.  Within the context of this limited
study we can make tentative conclusions about the the superconductivity
parameters of U(1) and SU(2).  In U(1) we see no violation of the
London theory which indicates that the flux tube is described very well
by the Ginzburg-Landau theory in the extreme type II limit, i.e. $\xi
\ll \lambda$.  In SU(2) we find these two quantities are of the same
order of magnitude.  The Ginzburg-Landau parameter $\kappa \equiv
\lambda/\xi = 0.7\pm0.1$.  A value of $\kappa$ less than $1/\sqrt{2}$
indicates that the superconductor is type II, otherwise type I.  Our
results place $\kappa$  on the borderline.  Maedan {\em et
al.\/}\cite{s} have come to the same conclusion from an effective
Lagrangian approach.

\section{Acknowledgments}
We wish to thank A. Kotanski, A. Kronfeld, M. Polikarpov, G. Schierholz,
T. Suzuki, R. Wensley and K. Yee for many fruitful discussions.


\end{document}